\documentstyle[12pt]{article}
\makeatletter
\newcommand{\doublespacing}{\let\CS=\@currsize\renewcommand{\baselinestretch}
{1.5}\tiny\CS}
\newcommand{\lyxaddress}[1]{
  \par {\raggedright #1 
  \vspace{1.4em}
  \noindent\par}
}
\date{}
\begin{document}
%\begin{center}
%{\Large{\bf Entanglement teleportation via Bell mixture}}
\title{Entanglement teleportation via Bell Mixture}
%\vspace{0.5cm}
%{\bf Sibasish Ghosh}$^*$\footnote{E--mail address : res9603@www.isical.ac.in},
%{\bf Guruprasad Kar}$^*$, {\bf Anirban Roy}$^*$\footnote{E--mail address :
%res9708@www.isical.ac.in}, {\bf Debasis Sarkar}$^{**}$, \\ {\bf Ujjwal
%Sen}$^!$\footnote{E--mail address : dhom@bosemain.boseinst.ernet.in}
\author{Sibasish Ghosh\protect\( ^{1}\protect \)\ \thanks{res9603@isical.ac.in},
Guruprasad Kar\protect\( ^{1}\protect \)\ \thanks{gkar@isical.ac.in}, Anirban Roy\protect\( ^{1}\protect \)\ \thanks{res9708@isical.ac.in}, 
Debasis Sarkar\protect\( ^{2}\protect \)\ \thanks{dsarkar@cubmb.ernet.in},  \\ and Ujjwal sen\protect\( ^{3}\protect \)\ \thanks{ujjwalsen@yahoo.co.in}}
\maketitle
\noindent{
\lyxaddress{\protect\( ^{1}\protect \)Physics and Applied Mathematics Unit, Indian Statistical
Institute, 203 B. T. Road, Calcutta 700 035, India}}

\noindent{
\lyxaddress{\protect\(^{2}\protect \)Department of Applied Mathematics, University 
of Calcutta, 92 A. P. C. Road, Calcutta 700 009, India}}

\noindent{
\lyxaddress{\protect\( ^{3}\protect \)Department of Physics, Bose Institute, 93/1 A.
P. C. Road, Calcutta 700 009, India}}

%\vspace{0.4cm}
%{\bf Abstract} 
%\end{center} 
%\vspace{0.2cm}

\begin{abstract}
We investigate the teleportation of the bipartite entangled states through two equally noisy 
quantum channels, namely mixture of Bell states. There is a particular mixed state channel for
which all pure entanglement in a known Schmidt basis remain entangled after teleportation and it happens
till the channel state remains entangled. Werner state channel lacks both these features.  
The relation of these noisy channels with violation of Bell's inequality and 2-E inequality is
studied.
%\end{center}
\end{abstract}
%\newpage
\doublespacing
\section{Introduction}
Quantum teleportation is a well known phenomenon, proposed by Bennett {\it et
al.} (BBCJPW) \cite{BBCJPW}, in which an unknown state
can be sent exactly, without being physically transported, 
 to a distant party by local operations and clasical communication.
For this purpose an entangled state (in fact a maximally entangled state) 
is required as teleportation channel between the sender and the receiver \cite{lo}.
% For nonmaximally entangled channel teleportation fidelity have
%been studied and even for some mixed entangled channel teleportation fidelity 
%remains higher than that for classical channel.
Quite recently, transfer of 
entanglement has been studied in the context of quantum teleportation of an
unknown {\it entangled} state through two equally noisy quantum channels \cite{lee}.
Here by teleportation of entanglement we mean after teleportation the final two qubit state 
remains entangled, whatever the reduced density matrices of the final two qubit state be.  
An arbitrary two-qubit state ${\rho}_{12}$ was taken as the unknown entangled
state which was to be sent, perhaps inexactly, to $B_1B_2$ where there were two
noisy channels, one between $A_1$ and $B_1$ and another between $A_2$ and
$B_2$. 
%(fig.1). 
The same entangled state, namely  
%$${\rho}_W = ,$$
%\noindent{
the Werner state \cite{werner}, was taken for both channels and it was shown
that entanglement of the unknown state can be completely lost during the 
teleportation even
when the channel is quantum correlated. At this point one should note the
point that the Werner channel would stop
teleporting entanglement depends solely
on the maximally entangled fraction (MEF) \cite{note} of the channel and irrelevant of 
whether the entangled state to be teleported is known or unknown. For a 
given Werner channel, it starts teleporting pure entanglement (i.e final state remain entangled 
after being teleported) if it is sufficiently entangled i.e 
MEF of
the Werner state is greater than $1/\sqrt{3}$ ($MEF > 1/\sqrt{3}$). There is no Werner state channel which can
teleport all the entanglement known or unknown.
In this regard we search for some mixed channel which
can teleport all {\it pure} entanglement where Schmidt basis of states is known. Here by ``all entanglement", we mean a
class of states whose entanglement ranges continuously from $0$ to $1$, with
respect to some measure.
%}

In this paper, we study the teleportation of pure entanglement in the above sense where Bell
mixtures are considered as cahnnel state. We provide sufficient conditions, the channel state
has to satisfy for teleporting at least one pure entanglement as well as full range of pure entanglement.
 We also study how this sufficient condition is related to
violation of Bell's inequality \cite{horo1} and 2-E inequality \cite{horo2}.
 it is shown that there really
exist such mixed state channels (mixture of two Bell states) which can teleport 
all {\it pure} entanglement from the plane spanned by an arbitrary but fixed
Schmidt basis. 
Another important feature of this channel is that,unlike the Werner channel, 
 it can teleport 
entanglement as long
as it itself remains entangled.

\section{Bennett protocol and mixture of Bell states}
Consider the following situation. $A_1$ and $B_1$ are sharing a state which
 is a
mixture of two Bell states 
$$wP[|\Psi^{+}\rangle]+(1-w)P[|\Psi^{-}\rangle],$$
\noindent{where $|\Psi^{\pm}\rangle=(1/\sqrt{2})(|01 \rangle \pm |10 \rangle)$ and 
$0 \le w \le1$. $A_2$ and $B_2$ are also sharing the same state. Now $A_1$, 
$A_2$ are 
sharing an entangled state of two particle 1 and 2 respectively, which they want to telport through these noisy
channels to $B_1$, $B_2$. Let the state shared by $A_1$, $A_2$ be $|\phi \rangle= \alpha|00
\rangle + \beta|11 \rangle$ (with $|\alpha|^2 + |\beta|^2 = 1$). Now $A_1$, $B_1$ follow
a teleportation 
protocol (from now on which we call as ${\cal P}_{|\Psi^{+} \rangle}$) and $A_1$ wants to
teleport the state of 1 to $B_1$ \cite{foot2} by that protocol. The same protocol is used by $A_2$ 
and $B_2$. 
%$A_i$ and $B_i$ perform Bennett protocol for
%exact teleportation of state sepearately as if the shared channel is $\Psi^{+}$
Finally the state shared between $B_1$, $B_2$ is given by} 
$$ \left[
\begin{array}{cccc}
|\alpha|2  & 0 & 0 & (2w-1)^2\alpha \beta^* \\
0 & 0 & 0 & 0 \\
0 & 0 & 0 & 0 \\
(2w-1)^2\alpha^*\beta & 0 & 0 & |\beta|^2
\end{array}
\right]$$
\noindent{which is inseparable whenever $w \ne 1/2$. Interestingly, only for $w=1/2$ the
channel itself becomes separable. Note that each of the reduced density
matrices of the final state is same as the corresponding ones of the initial state.
Thus the mapping  ${\tilde{V}}$ from Hilbert-Schmidt space of particle 1(2) to that of
particle $B_1$($B_2$) is given by\
${\tilde{V}} : I |\!\!\!\rightarrow I$, ${\tilde{V}} : \sigma_z |\!\!\!\rightarrow \sigma_z$.
We provide an example of a mixed state as channel which can teleport any pure 
entanglement where the Schmidt basis of such states is known, 
unless the channel itself becomes disentangled.} 

We remember that Werner channels lack both these features: firstly Werner
channels cannot teleport all entanglement and secondly 
it can teleport entanglement only
when it is sufficiently entangled (when $MEF > 1/\sqrt{3}$).           

It can be easily shown that all entanglement in a unknown Schmidt basis can not be
teleported through this kind of channel.

\section{Sufficient condition for teleportation of entanglement}
Next we consider the most general channel state between $A_i$ and $B_i$ ($i = 1,2$) for which
there exists a teleportation
protocol which maps the basis operators of Hilbert-Schmidt space of one particle
to the Hilbert-Schmidt space of another particle in the following way:
$${\tilde V}: I~ |\!\!\rightarrow I, ~ {\tilde V}: \sigma_i~ |\!\!\rightarrow \lambda_i \sigma_i ~ (i=x, y, z),$$ 
where $\lambda_i$'s are real numbers s.t. $|\lambda_i| \le 1$.
\noindent{Now we show that when 
${\lambda}^2_x+{\lambda}_y^2+{\lambda}_z^2 > 1$, the
channel can teleport at least one entangled state.
Consider the maximally entangled state
$(1/\sqrt{2})(|00\rangle+|11\rangle)$, $|0\rangle, |1\rangle$ being eigenstates of ${\sigma}_z$.
 After the teleportation,
the state becomes}
$$\frac{1}{4} 
\left[
\begin{array}{cccc}
1+{\lambda}^2_z & 0 & 0 & {\lambda}_x^2+{\lambda}_y^2 \\
0 & 1-{\lambda}^2_z & {\lambda}_x^2-{\lambda}_y^2 & 0 \\
0 & {\lambda}_x^2-{\lambda}_y^2 & 1-{\lambda}_z^2 & 0 \\
{\lambda}_x^2+{\lambda}_y^2  & 0 & 0 & 1+{\lambda}_z^2
\end{array}
\right]$$
\noindent{which remains
entangled if and only if}
$${\lambda}^2_x+{\lambda}_y^2+{\lambda}_z^2 > 1.$$ 
\noindent{Let us provide an example of a channel state and a teleportation protocol which realizes the map ${\tilde V}$.
If one takes the teleportation protocol to be ${\cal P}_{|\Psi^{+} \rangle}$ \cite{foot2} and takes the channel state as a mixture of four Bell states, then the map
${\tilde V}$ is realised through this teleportation process in the following way \cite{GKRSS}:
Let the channel state be $\rho_{\rm ch} = w_1 P[\Psi^{+}]
+ w_2 P[\Psi^{-}] + w_3 P[\Phi^{+}] + w_4 P[\phi^{-}]$, ($w_i \ge 0$, $i=1,2,3,4$ and $\sum_{i=1}^4 w_i = 1$)
and if one follows the above mentioned teleportation protocol, one realises the map ${\tilde V}$ with
$\lambda_x = w_1 - w_2 + w_3 - w_4, \lambda _y = w_1 - w_2 - w_3 + w_4, \lambda_z 
= w_1 + w_2 - w_3 - w_4$. 

It is clear that just nonzero entanglement of channel state $\rho_{\rm ch}$ does not
guarantee entanglement teleportation (see for example, ref. \cite{lee}). In the absence of 
any necessary condition on the channel state for teleporting pure entanglement, we inquire for some 
sufficient physical condition on the channel state, namely its relation with violation of Bell's
inequality \cite{horo1} and 2--E inequality \cite{horo2}. 

(1) The channel state (Bell Mixture) between $A_i$ and $B_i$ ($i=1,2$) can teleport at least one pure entanglement
if it violates Bell's inequality. Following Horodecki's result \cite{horo1} a two qubit state $\rho$ violates
Bell's inequality iff the sum of two greatest eigenvalues of ${T}_{\rho}^{T}T_{\rho}$ is greater than 1,
where $T_{\rho}$ is the T matrix of $\rho$ in the standard Hilbert-Schmidt representation \cite{horo1}.
 
Now the state $\rho_{\rm ch}$ violates Bell's inequality iff $${\rm max}[[\{2(w_1 + w_3) - 1\}^2
+ \{2(w_1 + w_4) - 1\}^2], [\{2(w_1 + w_3) - 1\}^2 + \{2(w_3 + w_4) - 1\}^2], $$
$$[{2(w_4 + w_3) - 1}^2
+ {2(w_1 + w_4) - 1}^2]] > 1,$$ i.e.
$${\rm max}[({\lambda}^{2}_{x} + {\lambda}^{2}_{y}), ({\lambda}^{2}_{z} + {\lambda}^{2}_{y}), 
({\lambda}^{2}_{x} + {\lambda}^{2}_{z})] > 1,$$ 
which is stronger than ${\lambda}^2_x+{\lambda}_y^2+{\lambda}_z^2 > 1$.

(2) The channel state can teleport some pure entangelement if it violates the 2-E inequality \cite{horo2}.

2-E inequality for bipartite density matrix is given by $$S_2(\rho_{12}) \ge {\rm max} \{S_2(\rho_1), S_2(\rho_2) \}$$
where $\rho_1$ and $\rho_2$ are the reduced density matrices of $\rho_{12}$ and
$$S_2(\rho) = - {\rm ln}{\rm Tr} \rho^2$$ 

One can easily check that the state $\rho_{\rm ch}$ violates the 2-E inequality \cite{horo2} 
iff $\lambda_x^2 + \lambda_y^2 + \lambda_z^2 > 1.$ Thus if the channel state $\rho_{\rm ch}$
 violates 2-E inequality at least one entanglement can be teleported through this channel.

One should note that for Werner state channel ({\it i.e.}, $w_2 = w_3 = w_4$), violation of 2-E
inequality is necessary and sufficient condition for teleporting at least one entanglement.

\section{Teleportation of All Pure Entanglement}
Now we consider the special case where any one of the $\lambda$'s say,           
$\lambda_z = 1$ and the rest, {\it i.e.}, $\lambda_x$, $\lambda_y$ $\ne 0$. 
%Then all pure 
%entanglement can
%be teleported through such channels. 
And we want to teleport the state $\alpha|00\rangle + \beta|11\rangle$ (where $\sigma_z |0\rangle = |0\rangle~
{\rm and}~ \sigma_z |1\rangle = -|1\rangle$ and $|\alpha|^2 + |\beta|^2 = 1$)
by using our teleportation
protocol ${\cal P}_{|\Psi^{+} \rangle}$, given by the corresponding map ${\tilde V}$. So after teleportation, this state becomes
$$ \left[
\begin{array}{cccc}
|\alpha|^2 & 0 & 0 & \frac{\alpha^*\beta({\lambda}_x - {\lambda}_y)^2 + \alpha\beta^*({\lambda}_x + {\lambda}_y)^2}{4} \\
0 & 0 & \frac{{\rm Re}\{\alpha\beta^*\}({\lambda}_x^2-{\lambda}_y^2)}{2} & 0 \\
0 & \frac{{\rm Re}\{\alpha\beta^*\}({\lambda}_x^2-{\lambda}_y^2)}{2} & 0 & 0 \\
\frac{\alpha^*\beta({\lambda}_x + {\lambda}_y)^2 + \alpha\beta^*({\lambda}_x - {\lambda}_y)^2}{4} & 0 & 0 & |\beta|^2
\end{array}
\right]$$
\noindent{which is entangled for all nonzero $\alpha,~\beta$. We now show that 
a mixture of two Bell states is the only solution of the 
map ${\tilde V}$ in this special case.

Consider the channel state as 
$\rho_{\rm ch}=w_1P[\Psi^{+}]+w_2P[\Psi^{-}]+w_3P[\Phi^{+}]+w_4P[\Phi^{-}]$
where $w_1+w_2+w_3+w_4=1$ and $w_1, w_2, w_3, w_4 \ge 0$. If we use the protocol ${\cal P}_{|\Psi^{+} \rangle}$,
then the mapping for $\sigma_z$ is $\sigma_z |\!\!\rightarrow (w_1+w_2-w_3-w_4)
\sigma_z$. So $\lambda_z = w_1+w_2-w_3-w_4 = 1$ gives us $w_1+w_2=1$ and $w_3 = w_4 = 0$. Again, for the same channel $\lambda_x (=
w_1-w_2-w_3+w_4)$ and $\lambda_y (= w_1-w_2+w_3-w_4)$, both has to be nonzero. And so
$w_1 \ne w_2$. Hence mixtures of two Bell states with unequal weights are the
only solutions. The other feature to be noted is that after teleportation through this type of
channel, the marginal density matrices of the pure states of type $\alpha |00
\rangle + \beta |11 \rangle$, remain unchanged, whereas, in case of Werner
channel the marginal density matrices of no entangled state remain unchanged.}     
%between $A_i$ and $B_i$.

Thus a sufficient condition that a channel realising the map
($\tilde{V} : I |\!\!\rightarrow I, \sigma_i |\!\!\rightarrow \lambda_i \sigma_i (i = x, y, z)$) can teleport
all pure entanglement (in a known basis) is that one of $\lambda_i$'s, say
$\lambda_z= 1$ and $\lambda_x, \lambda_y \ne 0$.
\section{Discussion}
In this paper, we investigated the effects of noisy channels on the
entanglement in entanglement teleportation. Here we have shown that the channel
composed of mixture of two Bell states with unequal weight can teleport a full
range of pure entanglement (although non-universally, i.e in a fixed basis), whereas the Werner
channel can not teleport all such entanglements. 
In this respect one can easily
see that to teleport any pure entanglement, the channel should violate 2-E
inequality.   
%We have shown that to transfer a
%full range (0 to 1) of entanglement, the nature of the noisy channel should be
%such that  
%$$I \rightarrow I, ~ \sigma_i \rightarrow \lambda_i \sigma_i ~
%(i=x, y, z).$$
%\noindent{where any of the $\lambda$'s is 1 and the other $\lambda$'s nonzero.
%An example of such a noisy channel is the mixture of two Bell states.  

\section{Acknowledgements}
U. S. acknowledges partial support
by the Council of Scientific and Industrial Research, Government of India, New
Delhi.


\begin{thebibliography}{99}
\bibitem{note} If the Werner state is $xP[\Psi^{-}] + \frac{(1-x)}{4}I$, then its MEF is $x$.
\bibitem{foot2} The protocol ${\cal P}_{|\Psi^{+} \rangle}$ is : Alice makes a projective measurement in Bell basis on her system, if she
gets $|\Psi^+\rangle$, then Bob does perform I on his system. Similarly for $|\Psi^-\rangle$,
unitary Operation be $\sigma_z$, for $|\Phi^+\rangle$ , it is $\sigma_x$ and for
$|\Phi^-\rangle$, $\sigma_z\sigma_x$
\end{thebibliography}
\end{document}